\documentclass[aps,prd,twocolumn,showpacs,superscriptaddress]{revtex4-1}

\usepackage{graphicx}  
\usepackage{bm}        
\usepackage{amssymb}   
\usepackage{amsmath} 

\graphicspath{{./figures/}}

\begin{document}


\title{Search for invisible modes of nucleon decay in water with the SNO\raisebox{0.35ex}{\small\textbf{+}} detector}

\author{M.\,Anderson}
\affiliation{\it Queen's University, Department of Physics, Engineering Physics \& Astronomy, Kingston, ON K7L 3N6, Canada}
\author{S.\,Andringa}
\affiliation{\it Laborat\'{o}rio de Instrumenta\c{c}\~{a}o e  F\'{\i}sica Experimental de Part\'{\i}culas (LIP), Av. Prof. Gama Pinto, 2, 1649-003, Lisboa, Portugal}
\author{E.\,Arushanova}
\affiliation{\it Queen Mary, University of London, School of Physics and Astronomy,  327 Mile End Road, London, E1 4NS, UK}
\author{S.\,Asahi}
\affiliation{\it Queen's University, Department of Physics, Engineering Physics \& Astronomy, Kingston, ON K7L 3N6, Canada}
\author{M.\,Askins}
\affiliation{\it University of California, Berkeley, Department of Physics, CA 94720, Berkeley, USA}
\affiliation{\it Lawrence Berkeley National Laboratory, 1 Cyclotron Road, Berkeley, CA 94720-8153, USA}
\affiliation{\it University of California, Davis, 1 Shields Avenue, Davis, CA 95616, USA}
\author{D.\,J.\,Auty}
\affiliation{\it University of Alberta, Department of Physics, 4-181 CCIS,  Edmonton, AB T6G 2E1, Canada}

\author{A.\,R.\,Back}
\affiliation{\it Queen Mary, University of London, School of Physics and Astronomy,  327 Mile End Road, London, E1 4NS, UK}
\affiliation{\it University of Sussex, Physics \& Astronomy, Pevensey II, Falmer, Brighton, BN1 9QH, UK}
\author{Z.\,Barnard}
\affiliation{\it Laurentian University, Department of Physics, 935 Ramsey Lake Road, Sudbury, ON P3E 2C6, Canada}
\author{N.\,Barros}
\affiliation{\it University of Pennsylvania, Department of Physics \& Astronomy, 209 South 33rd Street, Philadelphia, PA 19104-6396, USA}
\affiliation{\it Technische Universit\"{a}t Dresden, Institut f\"{u}r Kern und Teilchenphysik, Zellescher Weg 19, Dresden, 01069, Germany}
\affiliation{\it Laborat\'{o}rio de Instrumenta\c{c}\~{a}o e  F\'{\i}sica Experimental de Part\'{\i}culas (LIP), Av. Prof. Gama Pinto, 2, 1649-003, Lisboa, Portugal}
\author{D.\,Bartlett}
\affiliation{\it Queen's University, Department of Physics, Engineering Physics \& Astronomy, Kingston, ON K7L 3N6, Canada}
% ...
\collaboration{The SNO\raisebox{0.5ex}{\tiny\textbf{+}} Collaboration}

\date{\today}

\begin{abstract}
  This paper reports results from a search for nucleon decay through
  ‘invisible’ modes, where no visible energy is directly deposited during the
  decay itself, during the initial water phase of
  SNO\raisebox{0.5ex}{\tiny\textbf{+}}.  However, such decays within the oxygen
  nucleus would produce an excited daughter that would subsequently de-excite,
  often emitting detectable gamma rays. A search for such gamma rays yields
  limits of 2.5$\times 10^{29}$~y at 90\% Bayesian credibility level (with a
  prior uniform in rate) for the partial lifetime of the neutron, and
  3.6$\times 10^{29}$~y for the partial lifetime of the proton, the latter a
  70\% improvement on the previous limit from SNO.  We also present partial
  lifetime limits for invisible dinucleon modes of 1.3$\times 10^{28}$~y for
  $nn$, 2.6$\times 10^{28}$~y for $pn$ and 4.7$\times 10^{28}$~y for $pp$, an
  improvement over existing limits by close to three orders of magnitude for
  the latter two.
\end{abstract}


\maketitle

\section{Introduction}

Violation of baryon number conservation is predicted by many Grand Unified
Theories~\cite{BaryonViolationGUTs} potentially explaining the
matter-antimatter asymmetry of the universe. Searches for the decay of
protons or bound neutrons act as important constraints on our understanding of
physics beyond the Standard Model.

Modes of nucleon decay involving visible energy deposition by decay products, such
as positrons, pions or kaons, have so far not been observed by large scale
detectors~\cite{KamiokandeNDecay,KamiokandeNDecay2,KamiokandeNDecay3}. As such, interest has
turned to less well-constrained decay modes that could have escaped detection. A potential set of
channels are the invisible decay modes in which one or more nucleon decays to final
states which go undetected, such as those with neutrinos, other exotic neutral,
weakly-interacting particles or charged particles with velocities below the
Cherenkov threshold. Although no prompt signal would be observed from the particles
directly emitted in the decay, the remaining nucleus would be left in an excited state,
and could then emit a detectable signal as it de-excites. The search for the
de-excitation signal of the final state nucleus is model-independent, as it puts
no requirements on the particles produced in the decay. Theoretical models include modes with decays to three
neutrinos~\cite{3nuDecay} or to non-Standard Model particles such as the
unparticle~\cite{unparticle} or dark fermions~\cite{darkfermions}, the latter
providing a possible solution to the neutron lifetime
puzzle~\cite{NeutronDecayAnomaly}.

The Sudbury Neutrino Observatory (SNO) and KamLAND experiments have
conducted searches for such model-independent modes with KamLAND setting the
current best limit for the invisible neutron decay lifetime of $\tau(n\rightarrow
\mathrm{inv.})$~\textgreater~$5.8 \times 10^{29}$~y at 90\%
C.L.~\cite{KamLANDNucleonDecay} and SNO setting the best limit for invisible proton
decays of $2.1 \times 10^{29}$~y~\cite{SNONucleonDecay},  improving on previous
limits by the Borexino Counting Test Facility (CTF)~\cite{Back:2003wj} and
Kamiokande~\cite{KamiokandeNvvv}. Limits also exist for the dinucleon modes of
1.4$\times 10^{30}$~y~\cite{KamLANDNucleonDecay} for the $nn$ mode from
KamLAND, 5.0$\times 10^{25}$~y~\cite{Back:2003wj} for the $pp$ mode by the
Borexino CTF and 2.1$\times 10^{25}$~y~\cite{Tretyak:2004ze} for the $pn$ mode.

The SNO\raisebox{0.5ex}{\tiny\textbf{+}} experiment has been running since
December 2016, taking commissioning data with the detector filled with
ultrapure water. During this phase, a new search has been made for
invisible nucleon decay via the decay of $^{16}$O. After an
invisible nucleon decay, the $^{16}$O is left in either the $^{15}$O* excited
state, if the decaying nucleon was a neutron, or in the $^{15}$N* state, if the
decaying nucleon was a proton.  44\% of the time, $^{15}$O* will de-excite
to produce a 6.18~MeV gamma, and 2\% of the time, the decay will produce a 7.03~MeV
gamma. Similarly, $^{15}$N* will produce a 6.32~MeV gamma in 41\% of
decays, with 7.01, 7.03 and 9.93~MeV gammas produced 2, 2 and 3\% of the time,
respectively~\cite{EjiriO16}.
\begin{figure*}[]
  \centering
  \includegraphics[width=0.48\textwidth]{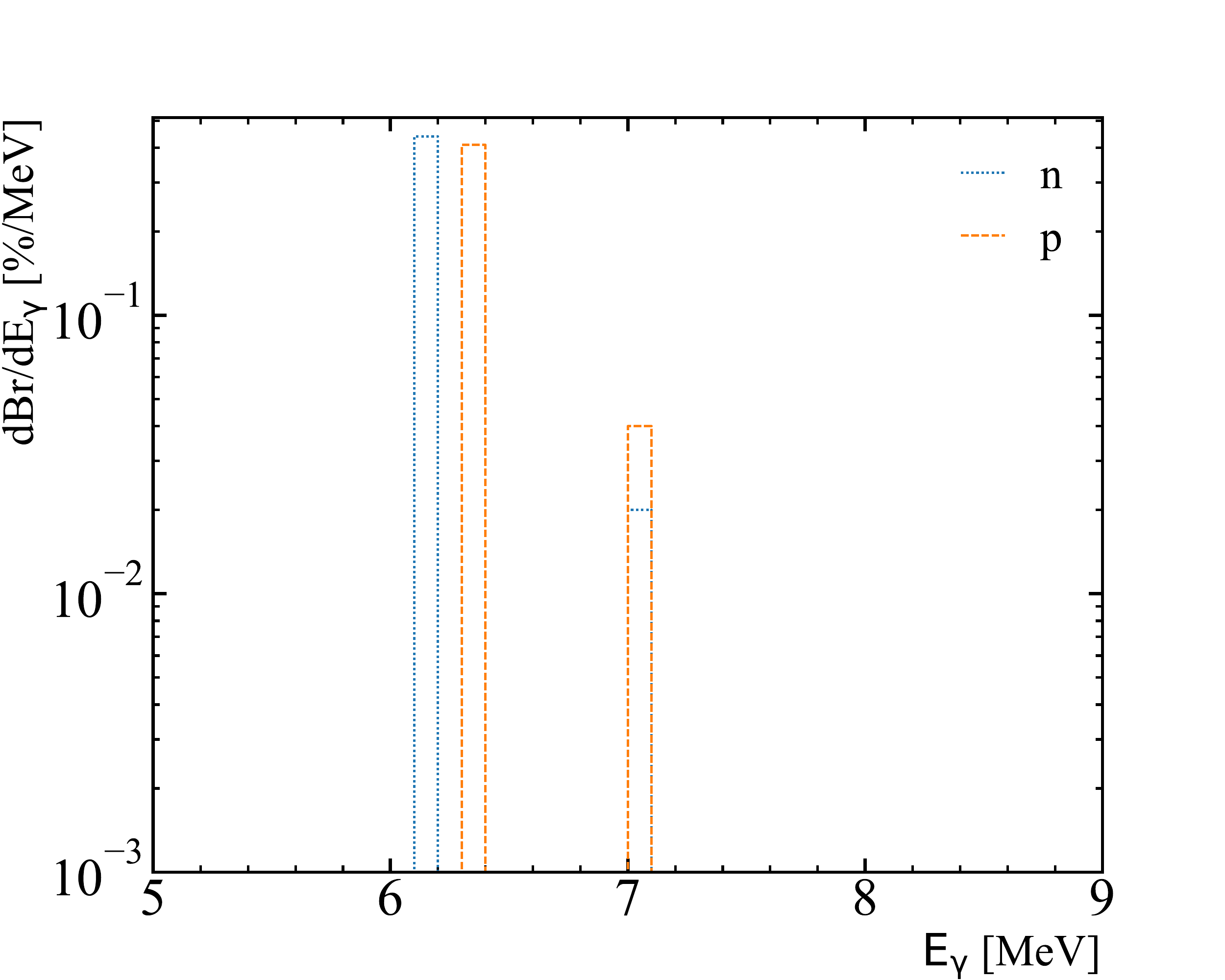}
  \includegraphics[width=0.48\textwidth]{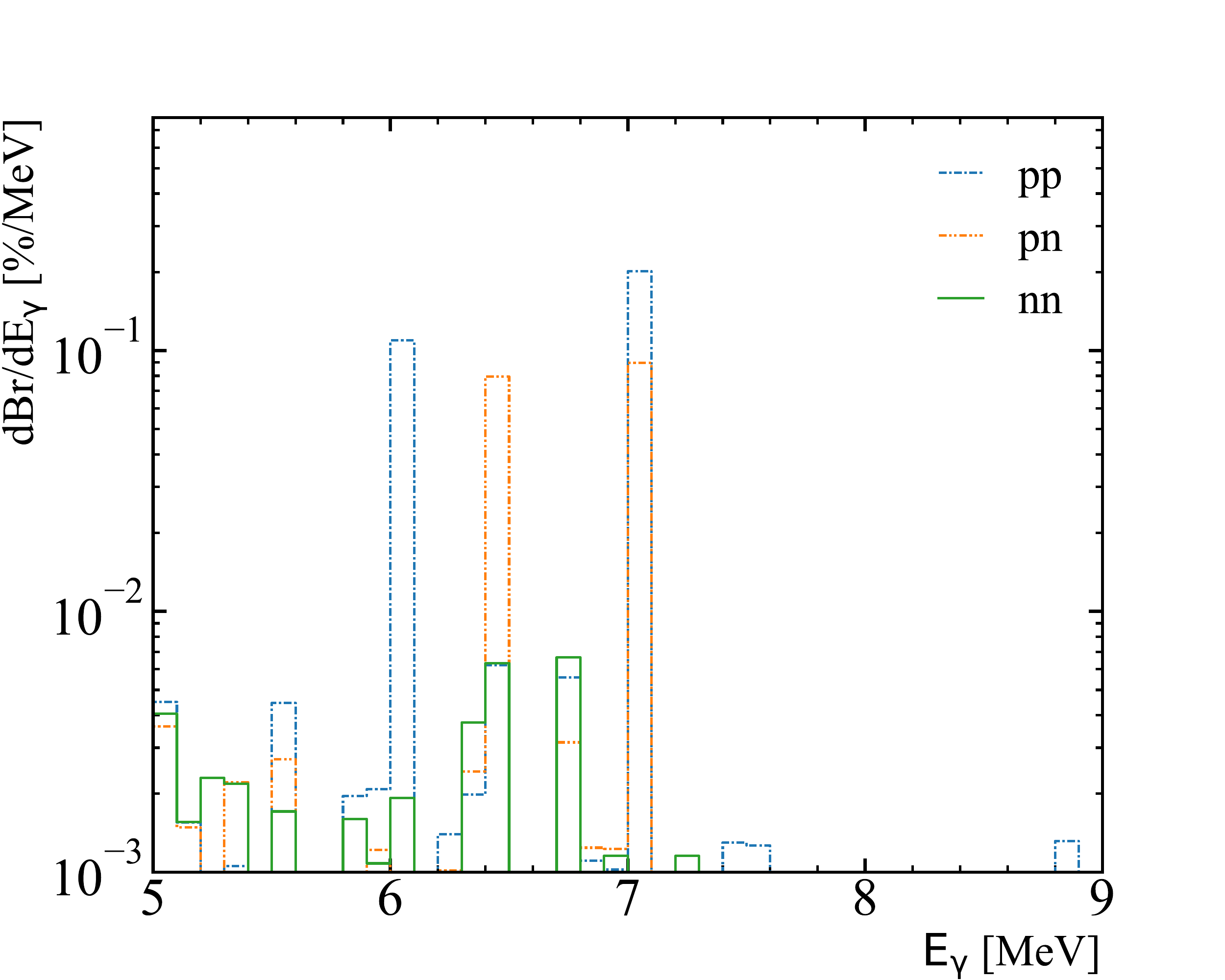}
  \caption{Spectrum of de-excitation gamma rays emitted after the invisible
  decay of $^{16}$O for single nucleon decay~\cite{EjiriO16} (left) and
  dinucleon decay~\cite{HaginoNirkko} (right) as a function of gamma ray energy
  $E_\gamma$ between 5 and 9 MeV.}
  \label{fig:decay_gammas_double}
\end{figure*}

This search has a unique sensitivity for two reasons: firstly, the branching
fraction to produce a visible signal of a de-exciting oxygen nucleus is larger
than the 5.8\% for carbon~\cite{KamyshkovC12} used by KamLAND. Secondly, the
use of H$_{2}$O rather than heavy-water (D$_{2}$O) removes the solar neutrino charged-current and neutral-current signals, major backgrounds in the SNO search.  

Dinucleon modes are also sought, based on the emission of de-excitation gamma
rays from $^{14}$O$^*$, $^{14}$N$^*$ and $^{14}$C$^*$ for the $nn$,
$pn$ and $pp$ invisible decay modes, respectively~\cite{HaginoNirkko}. The $pp$
decay can lead to gammas of 6.09~MeV at 10.9\% and 7.01~MeV at 20.1\%, and the
$pn$ decay has a 6.45~MeV gamma with 7.7\% and 7.03~MeV gamma with 8.9\%
probability. The $nn$ decay proceeds via many channels, with a summed
branching ratio of 4.53\% for gamma emission between 5 and 9 MeV. The branching
ratio for single and dinucleon decay are shown in Fig.~\ref{fig:decay_gammas_double}.

\section{The SNO\raisebox{0.5ex}{\tiny\textbf{+}} Detector}
The SNO\raisebox{0.5ex}{\tiny\textbf{+}} detector is inherited from SNO~\cite{SNODetector}, with several major upgrades to enable the use of liquid scintillator as the primary target rather than D$_{2}$O. The
detector consists of a 6~m-radius spherical acrylic vessel (AV) surrounded by 9394
inward-facing photomultiplier tubes (PMTs) mounted on a stainless steel support
structure with an 8~m radius from the center of the AV. During the initial water
phase, the AV was filled with approximately 905~tonnes of ultrapure water. The
cavity where the detector is installed is also filled with ultrapure water
to shield the innermost regions from radioactivity in the PMTs and the
surrounding rock.
Among the upgrades for SNO\raisebox{0.5ex}{\tiny\textbf{+}} was a new rope net to counteract the buoyancy of the
scintillator and hold down the AV~\cite{SNOPlusRopes}. The PMTs and front-end
electronics have been reused, with work done to repair PMTs that
failed over the lifetime of SNO. Aspects of the trigger system and data
acquisition (DAQ) software were upgraded to handle the higher data rates and
light yield expected in the scintillator phase.

\section{Data Selection}
\label{sec:DataSelection}
The results reported in this paper are based on the analysis of 235~days of
data recorded between May 4th 2017 and December 25th 2017. During this time,
the detector was live for 95\% of the time, with 16.9\% of that spent
performing calibration or maintenance.  A series of data quality checks were 
made to select the data for analysis with specific selection criteria for the
detector state, event rate, occupancy, and number of poorly calibrated channels,
resulting in the rejection of 29.3\% of the live time. The removal of time-correlated instrumental effects, cosmic ray muon events, and trigger dead time between events resulted in the loss of an additional 2.4\% of live time. The final
analyzed data set has a live time of 114.7~days with an uncertainty of
0.04\%. 

During the SNO\raisebox{0.5ex}{\tiny\textbf{+}} water phase, significant work was done on commissioning the
water processing and recirculation systems.  Changing background levels
associated with these activities motivated a time-dependent analysis.  The data
were split into six data sets, during each of which the background levels were
relatively stable, each with its own background estimate and set of analysis
cuts.  Table~\ref{tbl:livetime_timebins} details the live time of each data
set.

\begin{table}
\begin{tabular}{c|c|c|c|c|c|c|c}
\hline \hline
Data set   & 1 & 2 & 3 & 4 & 5 & 6 & Total \\
\hline
Duration (d) & 5.1 & 14.9 & 30.7 & 29.4 &  11.5 & 23.2 & 114.7\\
\hline \hline
\end{tabular}
\caption[ND decay time bins]{Live time within each data set, in days.}
\label{tbl:livetime_timebins}
\end{table}

\label{sec:PMTSelection}
Channels that failed calibration checks were excluded from the analysis,
though they still contributed to the hardware trigger.
The number of offline channels varied over time but
on average was around 800 channels.  A stable and well-calibrated channel can
still register hits caused by electronic cross-talk and PMT dark noise.
Hit-cleaning algorithms, used to exclude cross-talk hits from the analysis,
typically remove approximately 2\% of hits in an event. The dark noise is measured and
simulated on a run-by-run basis.

\section{Event Reconstruction}
\subsection{Vertex reconstruction}
\label{sec:recon}

SNO\raisebox{0.5ex}{\tiny\textbf{+}} uses a set of algorithms to reconstruct the position and direction of
Cherenkov events based upon maximizing the likelihood of a distribution of PMT
hit times that have been corrected for the time residuals, i.e., the
time of flight relative to the assumed vertex position, and of the angle
between the true direction and the line from the trial position to the PMT.
These algorithms only consider hits on well-calibrated online channels,
within 50~ns of the modal PMT hit time.

Three additional observables were used to classify events. The In Time Ratio (ITR)
is the ratio of PMT hit time residuals within a prompt timing window of [$-2.5$~ns, 5~ns]
to the total PMT  hit time residuals in an event. $\beta_{14}$ is an isotropy parameterization based on the 1st
and 4th Legendre polynomials of the distribution of PMT hits in the
event~\cite{LETA}, calculated using the angle between two PMTs
with respect to the reconstructed position. The projection of a particle's
reconstructed direction unit vector onto the corresponding event position unit
vector ${\bold u}\cdot {\bold r}$ determines whether the particle appears
inward- or outward-going relative to the center of the AV. For the physics and
calibration analyses, it is required that ITR$>$0.55 and
$-0.12<$$\beta_{14}$$<$0.95.

\subsection{Vertex systematic uncertainties}

Systematic uncertainties in the reconstructed vertex were evaluated using a
$^{16}$N calibration source~\cite{Dragowsky:2001ax}, previously used in the SNO
experiment, to produce tagged 6.1~MeV gammas.

A series of $^{16}$N scans were taken during the data taking period in 2017.
During a scan, $^{16}$N data was collected in a series of runs at points spaced about 50~cm apart
along the principal axes of the detector, typically through the center along the
x, y and z axes, where the z axis points upwards through the neck of the AV.
Additional scans were also taken off-axis in the xz and yz planes to evaluate
asymmetries in the detector and reconstruction.

\subsubsection{Position uncertainties}
\begin{figure}
	\centering
	\includegraphics[width=0.48\textwidth]{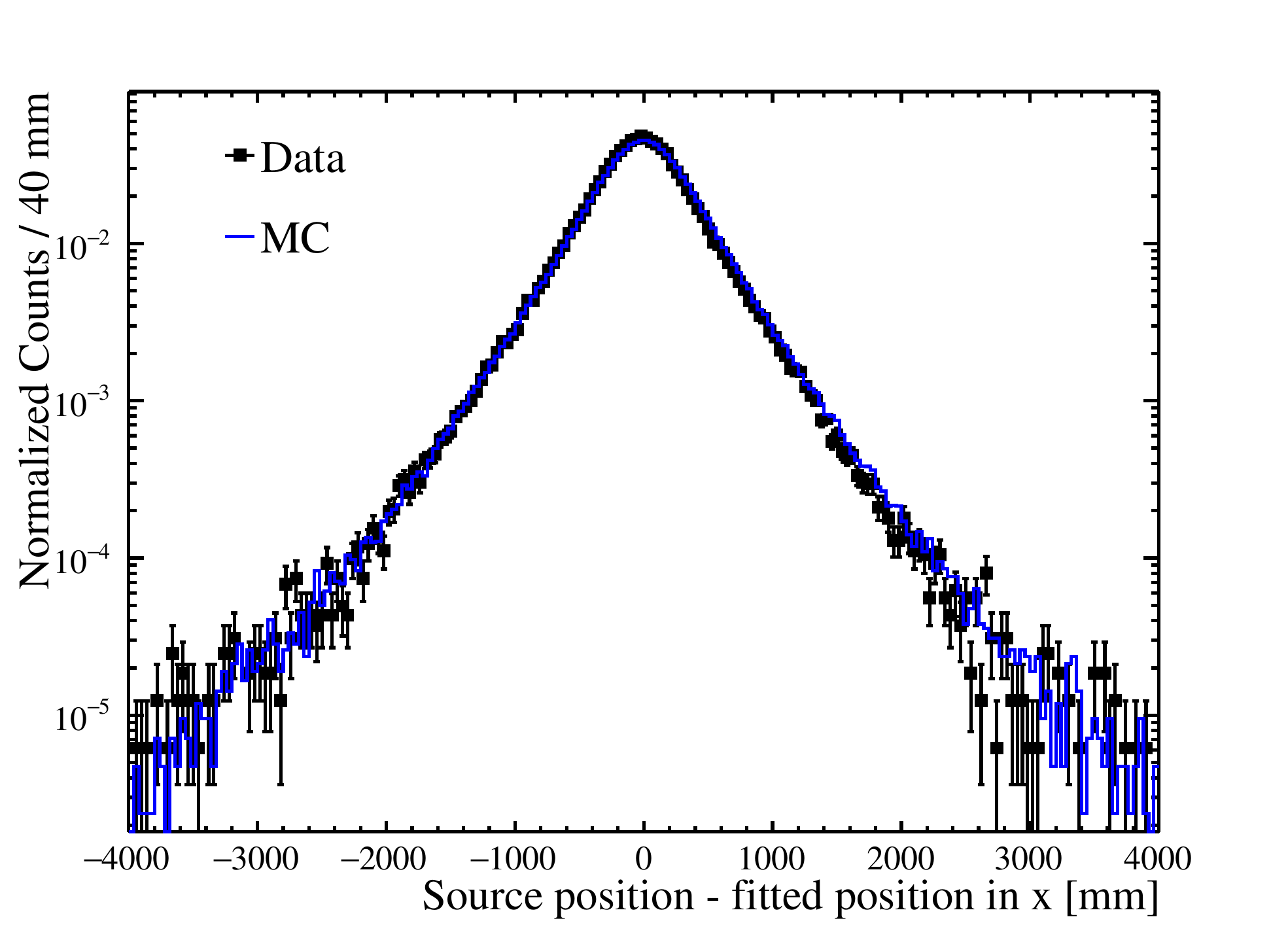}
        \caption{Comparison of the x-component of reconstructed position
        between $^{16}$N data acquired at a central source position and Monte
        Carlo simulation.} 
        \label{fig:ReconPositionFit}
\end{figure}

To evaluate uncertainties associated with the reconstructed event vertex
position and direction, the measured detector response to the $^{16}$N
calibration source was compared with predictions from Monte Carlo simulation, shown in
Fig.~\ref{fig:ReconPositionFit}.

For events which were tagged by the source PMT and passed the $\beta_{14}$ and ITR
cuts, the difference in the reconstructed vertex position and source position
was taken in each of the x, y and z axes. The resulting one-dimensional
distributions were fit with a distribution function representing the position
of the first Compton electron, estimated from the Monte Carlo model, convolved
with a Gaussian function and an exponential tail.

The uncertainties in reconstruction were characterized in terms of a constant
offset between the position of the source and the mean reconstructed position, 
x$\rightarrow \mathrm{x} + \delta_i$, a position-dependent scale
factor in which the position offset scales linearly with its value,
$\mathrm{x}\rightarrow (1 + \frac{\delta_i}{100})\mathrm{x}$, and a resolution
describing the width of the distribution of reconstructed event positions,
$\mathrm{x}\rightarrow \mathrm{x} + \mathcal{R}(0, \delta_i)$, where
$\mathcal{R}$ is a random number drawn from a Gaussian distribution of width
$\delta_i$ and mean 0.

\begin{table}[htbp]
\centering
\begin{tabular}{c | c}
\hline
\hline
Parameter & Uncertainty, $\delta_i$  \\ 
\hline
x offset (mm) & $^{+16.4}_{-18.2}$ \\
y offset (mm) & $^{+22.3}_{-19.2}$ \\
z offset (mm) & $^{+38.4}_{-16.7}$ \\
\hline                              
x scale (\%) & $^{+0.91}_{-1.01}$ \\
y scale (\%) & $^{+0.92}_{-1.02}$ \\
z scale (\%) & $^{+0.92}_{-0.99}$ \\
\hline
x resolution (mm) & 104 \\
y resolution (mm) & 98	\\
z resolution (mm) & 106	\\
\hline
Angular resolution & $^{+0.08}_{-0.13}$ \\
\hline
$\beta_{14}$		&	$\pm 0.004$ \\
\hline
\hline
\end{tabular}
\caption{Systematic uncertainties in the reconstructed position and direction of events based on analysis of $^{16}$N data.}
\label{tab:PosDirUncertanties}
\end{table}

The vertex offset was calculated as the volume-weighted mean of the difference
between the Gaussian fitted means of data and Monte Carlo simulation while the scale was found
as the slope of a linear fit to the differences, both listed in
Table~\ref{tab:PosDirUncertanties}. These were applied during signal extraction
by shifting and scaling the position of each event according to the
uncertainties along each axis independently and recomputing the
timing probability density functions (PDFs) used for signal extraction.

The position resolution of the data events was found to be $\sim$200~mm. The
difference in resolutions between the data and Monte Carlo events was modeled
as a Gaussian of standard deviation
$\delta_i=\sqrt{\sigma^2_\mathrm{Data} - \sigma^2_\mathrm{MC}}$
for each $^{16}$N run. The results were combined in a volume-weighted average,
independently for each detector axis. The resulting values for
$\delta_i$ are also listed in Table~\ref{tab:PosDirUncertanties}.
These were applied during the signal extraction, smearing the positions of all
Monte Carlo events by a Gaussian distribution of the appropriate width to
reproduce the data.

\subsubsection{Angular resolution}
Reconstructed direction is also evaluated using the $^{16}$N source, by taking
into account the high degree of colinearity between Compton scattered electrons
and the initial gamma direction. The angle $\theta$ between the `true'
direction, taken to be the vector from the source position to the reconstructed
event vertex, and the reconstructed event direction was calculated and the
distribution of these angles were compared for data and Monte Carlo events. To
reduce the effect of position reconstruction uncertainties, only events that
reconstructed more than 1200~mm from the source position were used. The
resulting distributions were fit with a functional form of two exponentials as
employed by SNO~\cite{PhysRevC.72.055502}: 
\begin{eqnarray}
\begin{aligned}
  R(\cos\theta) = ~
      & \alpha_m \beta_m \frac{\mathrm{e}^{\beta_m(\cos\theta - 1)}}
      {1 - \mathrm{e}^{-\beta_m}} \\
      &+ (1 - \alpha_m)\beta_s\frac{\mathrm{e}^{\beta_s(\cos\theta - 1)}}
      {1 - \mathrm{e}^{-\beta_s}}\mathrm{,}
\label{eq:ReconAngularExponentials}
\end{aligned}
\end{eqnarray}
where $\beta_m$ and $\beta_s$ are the slopes of the two exponential components,
parameterizing the distribution for small and large angles respectively,
and $\alpha_m$ is the fraction of the events following the exponential with
slope $\beta_m$. The volume-weighted average of the differences in $\beta$
was computed across all runs and the resulting value used as an estimate of the
uncertainty in angular resolution, transforming $\cos \theta \rightarrow 1 +
(\cos \theta -1)(1 + \delta_i)$. Events that were transformed beyond $-1$ were 
randomly assigned a value between $-1$ and 1. The resulting uncertainties are
listed in Table~\ref{tab:PosDirUncertanties}.

\subsubsection{$\beta_{14}$ uncertainties}
The systematic uncertainty of $\beta_{14}$, shown in Table~\ref{tab:PosDirUncertanties}, 
was determined from the difference between the
means of Gaussian fits to data and Monte Carlo simulations of a $^{16}$N run
with the source at the center of the detector.  This found a shift of $-0.031
\pm 0.004$, which was applied to the Monte Carlo predictions.

\subsection{Optical calibration}
The detector's optical parameters, including the attenuation and scattering of
light in the water and acrylic, and the PMT angular response, are based on
calibration measurements of the same materials from SNO~\cite{LETA}. Further
optical calibrations were carried out using the `laserball'~\cite{Laserball}, a multi-wavelength laser
pulse diffuser capable of running with 337, 365, 385, 420, 450 and 500~nm
wavelengths, deployed within the detector.

Using the laserball data, the group velocity of light in the SNO\raisebox{0.5ex}{\tiny\textbf{+}} water was
verified to be consistent with the values used in the SNO\raisebox{0.5ex}{\tiny\textbf{+}} simulation and
reconstruction~\cite{physConstants}.

Laserball runs were taken along the z axis after major detector maintenance
periods to re-measure the PMT gain, electronic time delays and their dependence
with integrated pulse charge. The delays were compared for the different
laserball runs, with time drifts typically less than 0.5~ns. Larger observed
drifts are consistent with actual changes in the detector hardware, for
example, the replacement of offline channels. 

From visual observation during detector upgrades, it is known that the
reflectivity of the PMT concentrators has degraded over time. The concentrator
diffuse reflectivity was tuned in simulation by comparing the PMT hit time
residual spectrum between a central $^{16}$N run and its Monte Carlo
simulation, with particular attention given to peaks in the late light (with
residual times between 47 and 80~ns) due to reflections from the concentrators, PMT glass,
and the AV. The total reflectivity was found to show no change but the diffuse
reflectivity was increased from 2.0\% to about 22\% to provide a better match
with the observed data.

The overall efficiency of the electronics and PMTs was calibrated by aligning
the simulated spectrum of the number of prompt PMT hits to that from the $^{16}$N calibration data at the center of the detector. 

\subsection{Energy reconstruction}
The kinetic energy of an event $T_e$ is reconstructed by comparing the
observed and expected numbers of prompt hits, defined as those with time
residuals within [$-10$, 8] ns, based on simulation using the reconstructed position and direction.
Since events are reconstructed under the hypothesis of an incoming electron, the
observable energy of a gamma particle will reconstruct below its true energy due to
the effects of Compton scattering and Cherenkov threshold, such that a 6~MeV gamma reconstructs around
5~MeV. Based upon an early $^{16}$N calibration run, 6.4 prompt PMT hits are expected
per MeV of electron-equivalent deposited energy.

\subsection{Energy systematic uncertainties}
\begin{figure}
	\centering
	\includegraphics[width=0.48\textwidth]{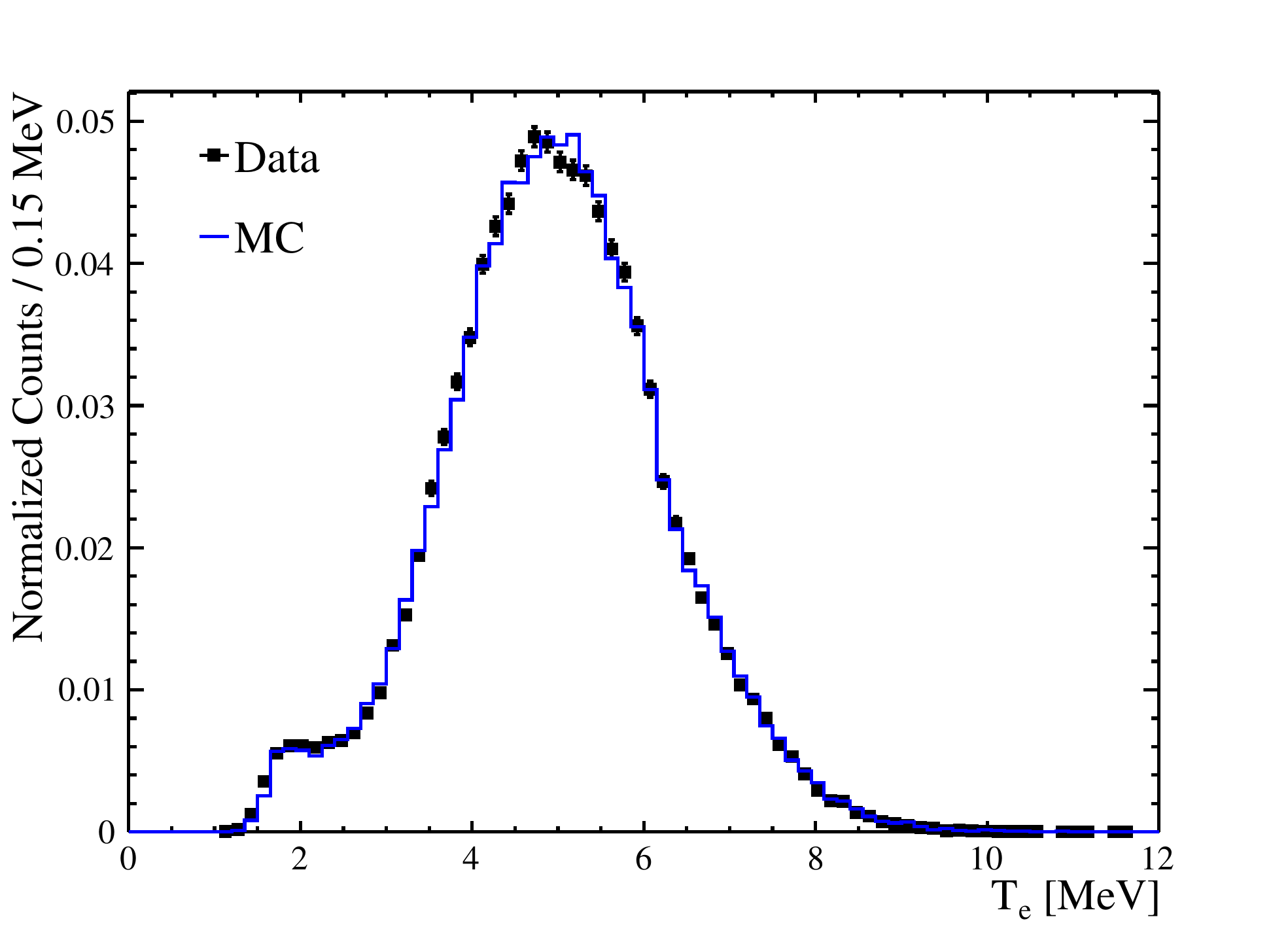}
        \caption{Comparison of reconstructed energy between $^{16}$N data
        acquired at a central source position and Monte Carlo simulation.}
        \label{fig:ReconEnergyFit}
\end{figure}

The relative energy scale and detector energy resolution were determined by
fitting the reconstructed energy spectrum from the $^{16}$N calibration source, as shown in
Fig.~\ref{fig:ReconEnergyFit},
with the predicted energy spectrum, generated from simulation, convolved with a
Gaussian distribution~\cite{Dunford:2006qb}. The fit is characterized in terms of a scale,
as a linear correction to the energy, and resolution, relating to the width of
the spectrum.

Events were associated with detector volume cells based on their reconstructed
positions.  The cells were distributed across z and $\rho \equiv
\sqrt{\mathrm{x}^2+\mathrm{y}^2}$, with dimensions 57~cm $\times$ 200~cm, determined based on statistics.
Events were also separated into seven bins based on their ${\bold u}\cdot
{\bold r}$ value. In addition to 82 distinct positions from scans within the AV,
the $^{16}$N source was also deployed at 19 positions along a vertical path between the AV and PMT array.  Since the mean free
path of the 6.1~MeV gamma is approximately 40~cm in water, all cells within the AV
contained data but, due to limitations in the deployment positions of the source, some cells outside the AV were not populated. The energy scale values of the cells, mapped in z and $\rho$, were fit with a continuous function to provide a smooth calibration of energy across the detector.  For each z-$\rho$ cell, fits were performed within all seven ${\bold u}\cdot {\bold r}$ bins and then averaged to provide a solid
angle-weighted energy scale in the cell. 

The deployment of the source was simulated at the same positions and with the same detector conditions. Half of each of the measured and simulated data sets were used to calibrate the energy scale as a function of position. The resulting calibrations were applied to the remaining halves of the $^{16}$N data sets, which were then fit to determine the relative energy scales and resolutions.
Uncertainties were volume and solid angle-weighted according to the selection criteria for different analysis regions. For the nucleon decay analysis, the relative energy scale and resolution uncertainties in the signal volume were 2.0\% and 1.8$\times\sqrt{\mathrm{T}_{e}/\mathrm{MeV}}\%$, respectively, where $\mathrm{T}_{e}$ is the reconstructed energy. The total energy uncertainties are dominated by the variation with position and the statistical uncertainty of the calibration data set. Contributions from source-related and time-related uncertainties were taken into account.  

\section{Background Model}

Several sources of background were considered for this search, mainly
from naturally occurring radioactive contamination from the $^{238}$U and
$^{232}$Th chains in the various detector components. Other sources include
interactions from solar neutrinos, atmospheric neutrinos and reactor neutrinos.
The analysis cuts, shown in Table~\ref{tbl:ndecay_box_cuts}, were chosen to
limit the impact of these backgrounds in the region of interest (ROI) for the
nucleon decay study.

\begin{table}
\centering
\begin{tabular}{c | c c c c c c c}
\hline
\hline
Data set & $\mathrm{T}_e$(MeV) & R (mm) & z (m) & $\cos\theta_{\odot}$\\
\hline
1     & 5.75--9 & $<$5450 & $<$4.0 & $<$0.80 \\
2 (z$>$0) & 5.95--9 & $<$4750 & $>$0.0  & $<$0.75 \\
2 (z$<$0) & 5.45--9 & $<$5050 & $<$0.0  & $<$0.75 \\
3     & 5.85--9 & $<$5300 & -    & $<$0.65 \\
4     & 5.95--9 & $<$5350 & $>-4.0$& $<$0.70 \\
5     & 5.85--9 & $<$5550 & $<$0.0 & $<$0.80 \\
6     & 6.35--9 & $<$5550 & -   & $<$0.70 \\
\hline
\hline
\end{tabular}
\caption{Summary of cuts that define the ROI used in the counting analysis for
  each data set, determined from Monte Carlo simulations based on expected
  background levels. Data set 2 is split into
  different cuts for the top and bottom halves of the detector as necessitated by backgrounds.
	Cuts of $-0.12<\beta_{14}<0.95$ and ITR $>0.55$ were used in every data set.
	No cuts were used on ${\bold u}\cdot {\bold r}$. The spectral analysis shares these cuts except for
  considering a broader energy range of 5-9~MeV.} 
  \label{tbl:ndecay_box_cuts}
\end{table}

\subsection{Instrumental backgrounds}
Backgrounds from detector instrumentation consisting of light emitted from
within the PMTs, electrical breakdowns in the PMT base or connectors, and
electronic pickup, were present during data taking.  A suite of data reduction
cuts were developed, based on those used by SNO, to remove these instrumental
backgrounds.  The cuts rely on low-level PMT information such as charge, hit
time, and hit location. The total
sacrifice within the nucleon decay ROI due to these cuts was measured to be 1.7\% by applying the
reduction cuts to $^{16}$N calibration source data. The remaining instrumental
contamination is evaluated using a bifurcated analysis method~\cite{LETA}, in
which two sets of data-cleaning cuts (bifurcation branches)
were used simultaneously on the analysis data; the instrumental contamination is
then calculated using the number of events that pass or fail one or both
bifurcation branches. Using the $^{16}$N sacrifice estimates, a correction to
the contamination estimate is also applied for the estimated signal events
flagged by the bifurcation branches. The number of events expected within each
data set are included in Table~\ref{tbl:ndecay_box_backgrounds}.

\subsection{Radioactive backgrounds}
\label{sec:backgrounds}

Three radioactive background analyses were performed in order to estimate the
contribution of $^{214}$Bi (U-chain) and $^{208}$Tl (Th-chain) decays from the
detector components and the detection medium in the nucleon decay ROI.
One analysis was dedicated to the radioactivity from U and Th chains in the the
water inside the AV, while two were dedicated to the radioactivity in the
acrylic itself, the hold-down rope system, the water shielding, and the PMTs.
Note that a new, sealed cover gas system in SNO\raisebox{0.5ex}{\tiny\textbf{+}}
to suppress radon ingress, from the headspace volume above and into the water
in the AV below, had not yet been brought online during the data taking period
reported here. This resulted in somewhat elevated and variable levels of
$^{214}$Bi in the AV water due to the lack of a radon-free cover gas.

\subsubsection{Internal radioactivity}\label{sec::bckg::internal}

\begin{table*}
\centering
\begin{tabular}{ c | c | c | c | c | c | c | c }
\hline
\hline
Data & \multicolumn{7}{c}{Expected events} \\
set  & Internal & External & Solar & Reactor & Atmospheric & Instrumental & PMTs \\
\hline
1 & 0.34$\pm0.04 ^{+1.25}_{-0.34}$ & 0.18$\pm0.03 ^{+0.48}_{-0.18}$ & 0.57$\pm0.01 ^{+0.03}_{-0.03}$ & 0.03$\pm0.00 ^{+0.01}_{-0.01}$ & 0.06$\pm0.00 ^{+0.04}_{-0.03}$ & 0.00$^{+0.06}_{-0.00}$ & 0.0$^{+4.6}_{-0.0}$ \\
2 & 0.70$\pm0.11 ^{+2.52}_{-0.70}$ & 0.39$\pm0.38 ^{+2.32}_{-0.39}$ & 1.05$\pm0.01 ^{+0.05}_{-0.07}$ & 0.08$\pm0.00 ^{+0.02}_{-0.02}$ & 0.13$\pm0.01 ^{+0.09}_{-0.07}$ & 0.00$^{+0.34}_{-0.00}$  & 0.0$^{+4.6}_{-0.0}$ \\
3 & 0.68$\pm0.09 ^{+2.83}_{-0.68}$ & 0.63$\pm0.12 ^{+1.27}_{-0.63}$ & 1.46$\pm0.02 ^{+0.08}_{-0.10}$ & 0.16$\pm0.00 ^{+0.03}_{-0.03}$ & 0.27$\pm0.02 ^{+0.18}_{-0.14}$ & 0.26$\pm0.17$ & 0.0$^{+4.6}_{-0.0}$ \\
4 & 0.91$\pm0.15 ^{+2.68}_{-0.91}$ & 0.42$\pm0.07 ^{+0.29}_{-0.32}$ & 1.57$\pm0.02 ^{+0.10}_{-0.11}$ & 0.10$\pm0.00 ^{+0.03}_{-0.02}$ & 0.25$\pm0.02 ^{+0.17}_{-0.13}$ & 0.13$\pm0.09$ & 0.0$^{+4.6}_{-0.0}$ \\
5 & 0.57$\pm0.12 ^{+2.14}_{-0.57}$ & 0.18$\pm0.04 ^{+0.39}_{-0.18}$ & 0.61$\pm0.01 ^{+0.04}_{-0.04}$ & 0.04$\pm0.00 ^{+0.01}_{-0.01}$ & 0.06$\pm0.01 ^{+0.04}_{-0.03}$ & 0.00$^{+0.07}_{-0.00}$  & 0.0$^{+4.6}_{-0.0}$ \\
6 & 0.58$\pm0.18 ^{+2.66}_{-0.58}$ & 0.17$\pm0.04 ^{+0.24}_{-0.17}$ & 1.18$\pm0.01 ^{+0.06}_{-0.07}$ & 0.08$\pm0.00 ^{+0.02}_{-0.02}$ & 0.15$\pm0.02 ^{+0.10}_{-0.08}$ & 0.08$\pm0.07$ & 3.6$^{+7.4}_{-2.3}$ \\\hline
\hline
\end{tabular}
\caption{The predicted number of events in the ROI passing counting analysis
  cuts for each data set, shown as the contribution from the water within AV
  (Internal), from the AV, ropes and water shielding (External) as well as solar, reactor
  and atmospheric neutrinos, instrumental backgrounds and backgrounds from the
  PMTs. The first uncertainty given is statistical, the second is the total
  systematic uncertainty. Instrumental backgrounds show only the systematic
	uncertainty while the backgrounds from the PMTs only include the statistical
	uncertainty.} 
  \label{tbl:ndecay_box_backgrounds}
\end{table*}

\begin{table*}[t]\centering
\scalebox{0.8}{
\begin{tabular}{c| c c| c c| c c| c}
\hline \hline
Period & \multicolumn{2}{c}{AV water} & \multicolumn{2}{c}{Water shielding} & \multicolumn{2}{c}{AV} & Ropes \\
& U & Th & U & Th & U & Th & Th \\
& [$\times 10^{-14}$ gU/g$_{H_{2}O}$] & [$\times 10^{-15}$ gTh/g$_{H_{2}O}$] & [$\times 10^{-13}$ gU/g$_{H_{2}O}$] & [$\times 10^{-14}$ gTh/g$_{H_{2}O}$] & [$\times 10^{-12}$ gU/g$_{AV}$] & [$\times 10^{-12}$ gTh/g$_{AV}$] & [$\times 10^{-9}$ gTh/g$_{rope}$] \\ \hline
1 & 19.0 $\pm$ 1.8 $^{+3.9}_{-3.7}$ & 5.9 $\pm$ 5.2 $^{+4.0}_{-5.9}$ & 2.2 $\pm$ 0.3 $^{+3.7}_{-1.3}$ & 9.9 $\pm$ 1.6 $^{+22.9}_{-9.7}$ & 5.5 $\pm$ 1.5 $^{+6.5}_{-5.5}$ & 0.0 $^{+0.0}_{-0.0}$ $^{+1.1}_{-0.0}$ & 0.0 $^{+0.0}_{-0.0}$ $^{+0.3}_{-0.0}$ \\[10pt]
2 (z$>$0) & 48.5 $\pm$ 3.1 $^{+11.7}_{-10.1}$ & 34.5 $\pm$ 13.7 $^{+11.2}_{-34.5}$ & 86.9 $\pm$ 1.1 $^{+103.2}_{-49.2}$ & 207.7 $\pm$ 6.4 $^{+449.9}_{-173.0}$ & 33.0 $\pm$ 16.4 $^{+60.8}_{-33.0}$ & 12.5 $\pm$ 2.4 $^{+33.9}_{-12.5}$ & 2.8 $\pm$ 0.5 $^{+7.7}_{-2.8}$ \\[10pt]
2 (z$<$0) & 3.6 $\pm$ 0.9 $^{+1.0}_{-0.7}$ & 2.7 $^{+4.2}_{-2.7}$ $^{+1.3}_{-2.7}$ & 16.3 $\pm$ 0.4 $^{+24.4}_{-8.5}$ & 39.8 $\pm$ 2.8 $^{+134.8}_{-39.8}$ & 7.7 $\pm$ 5.5 $^{+24.4}_{-7.7}$ & 3.7 $\pm$ 1.2 $^{+11.0}_{-3.7}$ & 0.9 $\pm$ 0.3 $^{+2.5}_{-0.9}$  \\[10pt]
3 & 8.7 $\pm$ 0.7 $^{+2.4}_{-1.7}$ & 8.3 $\pm$ 3.1 $^{+3.0}_{-8.3}$ & 1.7 $\pm$ 0.1 $^{+2.5}_{-1.1}$ & 9.3 $\pm$ 0.5 $^{+19.1}_{-9.1}$ & 1.2 $\pm$ 0.9 $^{+7.9}_{-1.2}$ & 0.0 $^{+0.3}_{-0.0}$ $^{+1.1}_{-0.0}$ & 0.0 $^{+0.1}_{-0.0}$ $^{+0.3}_{-0.0}$ \\[10pt]
4 & 19.4 $\pm$ 1.0 $^{+5.8}_{-4.4}$ & 9.4 $\pm$ 4.1 $^{+6.5}_{-9.4}$ & 0.6 $\pm$ 0.1 $^{+1.2}_{-0.4}$ & 10.6 $\pm$ 0.6 $^{+19.3}_{-8.8}$ & 0.3 $^{+0.8}_{-0.3}$ $^{+2.2}_{-0.3}$ & 0.0 $^{+0.1}_{-0.0}$ $^{+0.5}_{-0.0}$ & 0.0 $^{+0.0}_{-0.0}$ $^{+0.1}_{-0.0}$ \\[10pt]
5 & 53.5 $\pm$ 3.7 $^{+19.5}_{-14.3}$ & 29.0 $\pm$ 17.1 $^{+24.7}_{-29.0}$ & 2.3 $\pm$ 0.2 $^{+5.3}_{-1.6}$ & 8.6 $\pm$ 1.3 $^{+31.9}_{-8.6}$ & 5.2 $\pm$ 0.9 $^{+6.7}_{-5.2}$ & 0.1 $^{+0.5}_{-0.1}$ $^{+0.3}_{-0.1}$ & 0.0 $^{+0.1}_{-0.0}$ $^{+0.1}_{-0.0}$ \\[10pt]
6 & 67.5 $\pm$ 2.1 $^{+26.3}_{-20.8}$ & 67.1 $\pm$ 10.0 $^{+38.7}_{-67.1}$ & 1.2 $\pm$ 0.1 $^{+2.4}_{-0.8}$ & 10.0 $\pm$ 0.7 $^{+28.8}_{-10.0}$ & 1.7 $\pm$ 0.9 $^{+3.8}_{-1.7}$ & 0.0 $^{+0.1}_{-0.0}$ $^{+1.0}_{-0.0}$ & 0.0 $^{+0.0}_{-0.0}$ $^{+0.2}_{-0.0}$ \\[10pt]
    \hline \hline
\end{tabular}}
  \caption{Measured U and Th contamination in the AV water, water
  shielding, AV and ropes in g/g$_{H_{2}O}$, g/g$_{AV}$ and g/g$_{rope}$,
  respectively. The values for the AV water come from the internal background
  analysis while the water shielding, AV and ropes numbers come from the
  external fit analysis, with the AV and ropes scaled together. For the external fit analysis, the statistical uncertainty is obtained from the fit, while the systematic uncertainty includes the contamination from PMT events.}
  \label{tbl:bckgs}
\end{table*}

\begin{figure}
	\centering
	\includegraphics[width=0.48\textwidth]{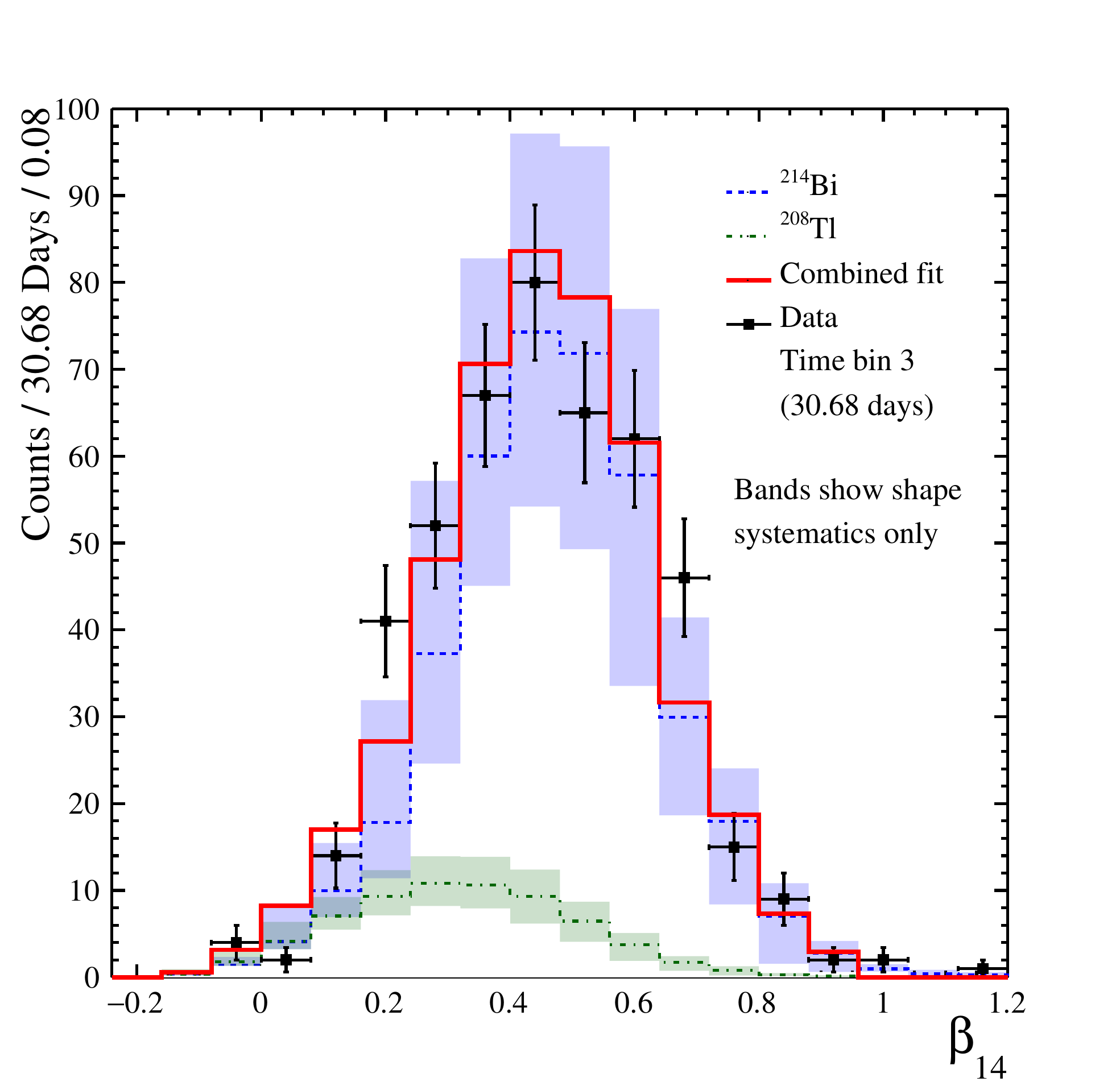}
        \caption{Data and Monte Carlo distributions of events in $\beta_{14}$ for $^{214}$Bi
        and $^{208}$Tl decays within the AV water analysis region for
        data set 3. The normalizations of the Monte Carlo spectra were fit to the
        observed data. }
	\label{fig:InternalsB14}
\end{figure}

$^{214}$Bi and $^{208}$Tl decays within the AV water were distinguished by
fitting to the $\beta_{14}$ distribution in a background analysis region defined by a radius of
4.3~m and energy above 4~MeV, to minimize the contamination from decays from
the AV and water shielding.  Monte Carlo simulations of $^{214}$Bi and
$^{208}$Tl decays were used to construct PDFs of the data within the background
analysis region.  In each of the data sets, the rates of $^{214}$Bi and $^{208}$Tl
were fit simultaneously to account for possible correlations, with the
resulting concentrations shown in Table~\ref{tbl:bckgs}. This is demonstrated
for data set 3 in Fig.~\ref{fig:InternalsB14}. The rate is then
extrapolated from the background analysis region into the nucleon decay ROI
based on the relative expected event rate between the two regions from MC simulations.

\subsubsection{External radioactivity}
Two independent analyses were performed to measure the radioactivity from the
AV, hold-down ropes and water shielding. A two-dimensional likelihood fit measured the rates of U and Th components based on $\beta_{14}$ as a function of R$^{3}$---where R is the reconstructed radius of the event as measured from the center of the AV---within a background analysis region chosen to preferentially contain events from the AV, ropes and water shielding. Results from the fit analysis are shown in Table~\ref{tbl:bckgs}. $^{214}$Bi and $^{208}$Tl were fit simultaneously, taking
into account correlations between the two.
The second analysis counted the events within four analysis regions, defined by R$^{3}$ and ${\bold u}\cdot {\bold r}$ to contain events from the water within the AV, AV and ropes, water shielding and PMTs respectively. Monte Carlo
simulations were used to translate the number of events observed in each into
total rates and to extrapolate to the expected number of events in the ROI,
shown in Table~\ref{tbl:ndecay_box_backgrounds}. The analysis took into
account the correlation of events between the different regions. Bias testing
of the fits ensured that asymmetries in the spatial distribution of events were
properly handled. The analyses agree with each other within uncertainties.

\subsubsection{PMT backgrounds}

Events from radioactive decays in PMTs have a very low probability to
reconstruct within the nucleon decay ROI, but occur at a very high rate, making
it difficult to estimate their contribution to the total background rate using
only Monte Carlo simulations. A data-driven method was used instead to put a
constraint on the rate of PMT background events in the ROI. The $^{208}$Tl
decay produces a beta particle alongside a gamma cascade that can occasionally
be detected in the PMT itself. The vertex
reconstruction of PMT events is dominated by the Compton scatter of the gamma
while the signal from the electron will appear early and concentrated in a
single PMT. Events with early hits and high charge were tagged as PMT events,
with simulation studies showing a tagging efficiency of close to 30\,\%, while
the misidentification rate is 1.4\,\%. The number of tagged events was used to
estimate the number of expected events originating from the PMTs.

\subsubsection{(alpha, n) interactions}
Another potential source of background is the set of (alpha, n) reactions on $^{13}$C
atoms within the AV itself, induced by alpha particles from the decay of
$^{210}$Po, a daughter of $^{222}$Rn. In about 10\% of the cases, a high
energy gamma or electron-positron pair is produced together with the neutron,
which can act as a background for the nucleon decay search. Using predictions
based on the leaching coefficient of $^{210}$Po in water, temperature and the
surface contamination on the AV~\cite{SnoplusReview}, less than 600
(alpha, n) decays were expected during the period of data taking.  Monte Carlo
simulations of these events predict less than 1 event reconstructing in the
nucleon decay ROI during the whole period under analysis.

\subsection{Neutrino induced backgrounds}
A dominant background for the nucleon decay search is the elastic scattering of $^{8}$B solar neutrino
interactions. Such events were largely excluded by a cut on $\cos\theta_{\odot}$, the reconstructed
direction relative to the direction to the Sun. Monte Carlo simulations of $^8$B solar
neutrinos were constrained based on recent measurements from
Super-Kamiokande~\cite{SuperKSolar}, with oscillations applied using the
BS2005-OP solar model~\cite{BS2005-OP}.

Antineutrinos produced by nuclear reactors also contribute to the background.
The expected number of events is estimated based on Monte Carlo simulations
using the world reactor power data~\cite{international2016iaea} with oscillations applied based on a global
best fit~\cite{Capozzi:2016rtj}.

Atmospheric neutrino interactions can create backgrounds for the nucleon decay
search, particularly through neutral-current interactions with the oxygen
nuclei. The interactions can liberate a nucleon from the $^{16}$O atom, leaving
either $^{15}$N* or $^{15}$O*, identical to the nucleon decay signal. However,
a large fraction of these events can be tagged by looking for neutron followers appearing 
after the initial event. In order to estimate the size of this background,
GENIE~\cite{Andreopoulos:2009rq, Andreopoulos:2015wxa} was used to
simulate high-energy atmospheric neutrino interactions. The GENIE Monte Carlo was
verified with two studies. One study counted nucleon decay-like
events with neutron followers to probe the size of the neutral-current background, and a second compared the energy, time and multiplicity of Michel electron followers directly to data.
Both searches found good agreement between GENIE and data and the difference
between the two is used as part of the atmospheric background uncertainty
estimate.

Due to the location of SNO\raisebox{0.5ex}{\tiny\textbf{+}} at a depth of 2092~m, equivalent to close to 6000~m of water,
the rate of cosmic-ray muons entering the detector is approximately three per hour.
Spallation products from these cosmic-ray muons are removed by cutting all events within
20~s of a muon event, as was used during SNO. This was shown to reduce the remaining number of events from spallation products to less than one event during the data taking period~\cite{SNOPhase1}.

\section{Analysis Methods}
A blind analysis was carried out, removing events with the number of PMT hits approximately corresponding to between 5 and 15~MeV from the data set. After the
analyses were finalized, blindness constraints were lifted and the whole of the
data was made available for analysis.

The expected signal was simulated using Monte Carlo techniques to develop
analyses that maximize the signal acceptance while minimizing the effect of
backgrounds. The observables for each event used in the search were the reconstructed
kinetic energy $\mathrm{T}_e$, the cube of the position radius R$^3$, normalized by the
cube of the radius of the acrylic vessel R$^{3}_{AV}$, position on the z-axis of the detector z,
and direction relative to the solar direction $\cos\theta_{\odot}$, as
well as the event classifiers $\beta_{14}$, ITR, and ${\bold u}\cdot {\bold r}$.

Two analysis methods, a spectral analysis and, as a cross-check, a counting analysis, were used to set a
limit at 90\% C.I. (credibility interval) on the number of signal decays (with a prior uniform in the decay rate), $S_{90\%}$. This is then
converted into a lifetime on the invisible nucleon decay modes using

\begin{equation}
\tau >  \frac{N}{S_{90\%}}\mathrm{,}
\end{equation}

\noindent where $N$ is the number of targets. For the neutron and proton modes,
this is defined as the number of neutrons (or protons) in $^{16}$O atoms inside 
a 6~m radius in the AV, $2.41\times10^{32}$. The
difference between this 6~m radius and the fiducial radius for a particular
data set is accounted for in the selection efficiency of that mode. For the
dinucleon modes, $N$ is defined as the number of $^{16}$O atoms within the
same volume, $3.02\times10^{31}$.

To calculate the limit on the number of decays from the limit on the observed
signal, an acceptance efficiency is calculated for each mode as the product of
the theoretical branching ratios~\cite{EjiriO16, HaginoNirkko} and the
selection efficiency from cuts on the observables with the total shown in
Table~\ref{tbl:ndecay_box_efficiencies}.

\begin{table}
\centering
\begin{tabular}{c | c | c | c | c | c}
\hline
\hline
Data & \multicolumn{5}{c}{Signal efficiency (\%)}\\
set  & $n$ & $p$ & $pp$ & $pn$ & $nn$\\
\hline
1 & 9.1 $^{+1.2}_{-1.2}$ & 11.2 $^{+1.1}_{-1.1}$& 10.4 $^{+0.8}_{-0.8}$& 5.9 $^{+0.5}_{-0.5}$ & 1.48 $^{+0.06}_{-0.06}$ \\
2 & 7.5 $^{+0.9}_{-0.9}$ & 9.2 $^{+0.8}_{-0.9}$ & 8.4 $^{+0.6}_{-0.6}$ & 4.8 $^{+0.4}_{-0.4}$ & 1.19 $^{+0.04}_{-0.04}$ \\
3 & 7.4 $^{+1.1}_{-1.1}$ & 9.3 $^{+1.0}_{-1.0}$ & 8.8 $^{+0.7}_{-0.7}$ & 5.0 $^{+0.4}_{-0.4}$ & 1.24 $^{+0.05}_{-0.04}$ \\
4 & 7.0 $^{+1.1}_{-1.1}$ & 8.8 $^{+1.0}_{-1.0}$ & 8.3 $^{+0.7}_{-0.7}$ & 4.8 $^{+0.4}_{-0.5}$ & 1.19 $^{+0.05}_{-0.05}$ \\
5 & 3.7 $^{+0.7}_{-0.6}$ & 4.9 $^{+0.6}_{-0.6}$ & 5.4 $^{+0.4}_{-0.5}$ & 3.1 $^{+0.3}_{-0.3}$ & 0.77 $^{+0.03}_{-0.03}$ \\
6 & 5.2 $^{+1.0}_{-0.9}$ & 7.1 $^{+1.0}_{-0.9}$ & 7.1 $^{+0.7}_{-0.7}$ & 4.1 $^{+0.4}_{-0.4}$ & 1.09 $^{+0.04}_{-0.04}$ \\
\hline
\hline
\end{tabular}
\caption{The acceptance efficiency of the signal after applying counting
  analysis cuts (for both single nucleon and dinucleon decay modes) of each
  data set. The single nucleon decay efficiencies are given per nucleon while
  the dinucleon modes are given for the nucleus as a whole. Combined systematic
  uncertainties are given, statistical uncertainties were negligible.}
  \label{tbl:ndecay_box_efficiencies}
\end{table}

\subsection{Spectral analysis}
A spectral analysis was performed, fitting for the signal in the measured
distribution of the observables $\bm{\eta} = [\mathrm{T}_e, \mathrm{R}^3,
\cos\theta_{\odot}, \beta_{14}, {\bold u}\cdot {\bold r}]$, within the limits defined in Table~\ref{tbl:ndecay_box_cuts} but with energy considered over the range 5--9 MeV for all data sets.
The backgrounds in the fit included solar
neutrinos, reactor neutrinos and atmospheric neutrinos as well as radioactivity
from U and Th chain decays in the water, AV, ropes, water shielding and PMTs.
Probability distributions for the signal and backgrounds,
$\mathcal{P}_s(\bm{\eta}_i)$ and $\mathcal{P}_{b}(\bm{\eta}_i)$, were
generated using Monte Carlo simulations with constraints on the radioactive
backgrounds provided by the likelihood-fit external analysis.

To allow for the multiple data sets, the analysis simultaneously maximized the
sum of the log likelihoods of each individual data set $k$, as described by:

\begin{eqnarray}
\label{eqn:fullLogLikelihood}
\begin{aligned}
& -\ln\mathcal{L}(s,\bm{\beta}|\bm{\eta},\hat{\bm{\beta}},\bm{\sigma},t_k) =
  \\&  -\sum_{k}\sum_{i=1}^{n_{obs}}\ln\biggl\{s\epsilon_{s,k} \mathcal{P}_{s,k}(\bm{\eta}_{i})
    +\sum_{b}\beta_{b,k}\epsilon_{b,k}\mathcal{P}_{b,k}(\bm{\eta}_{i})\biggr\}t_k
\\&  + \sum_{k}\biggl(s\epsilon_{s,k}+\sum_b\beta_{b,k}\epsilon_{b,k}\biggr)t_k +
\sum_{k}\sum_b\frac{(\beta_{b,k}-\hat{\beta}_{b,k})^2}{2\sigma_{b,k}^2}\mathrm{,}
\end{aligned}
\end{eqnarray}
where $n_{obs}$ is the number of observed events in each data set,
$s$ is the signal decay rate, $\epsilon_{s,k}$ is the acceptance
efficiency of the signal in data set $k$, $\beta_b$ is the rate of background
component $b$ whose expectation $\hat{\beta}_b$ is constrained by $\sigma_b$,
$\epsilon_b$ is the acceptance efficiency for the background $b$ and $t_k$ is
the live time of data set $k$. Fits were bias tested using a sampling
of fake data sets based on Monte Carlo simulations.

To find $S_{90\%}$, a profile likelihood~\cite{PDG2018} distribution is calculated by taking
the value of the maximum likelihood for a given value of $s$. The upper limit
at 90\% C.I. is then found by integrating along this distribution.

\subsection{Counting analysis}

A counting experiment with a set of rigid cuts, shown
in Table~\ref{tbl:ndecay_box_cuts}, is also used, where the number of background events
is calculated directly from the background analyses and is
shown in Table~\ref{tbl:ndecay_box_backgrounds}. Due to changes in the level of
backgrounds, candidate events were selected using different cuts during
different periods of running. The signal acceptance within each data set is
shown in Table~\ref{tbl:ndecay_box_efficiencies}. Using a Bayesian
method~\cite{OHelene}, an upper limit on the number of signal decays that could
have occurred is found by numerically solving

\begin{equation}
\int\limits^{S_{90\%}}_{0} \prod_{k} A \biggl(s\epsilon_{k}t_{k} + b_{k}\biggr)^{n_{k}} 
  \times \mathrm{e}^{-\left(s\epsilon_{k}t_{i}+b_{k}\right)} \mathrm{d}s/n_{k}! = 0.9\mathrm{,}
\label{eqn:NDecayTimebins}
\end{equation}

\noindent where $S_{90\%}$ is the upper limit on the number of signal decays at 90\%
credibility level and, for each data set $k$, $b_{k}$ is the number of expected
background events, combined from internal and external radioactivity, solar,
reactor, atmospheric and instrumental backgrounds, $n_{k}$ is the number of
observed events while $\epsilon_{k}$ and $t_{k}$ are the signal efficiency
after cuts and the live time of the data set. $A$ is a normalization factor such that the integral
tends to 1 as $S_{90\%}$ tends to infinity.

\section{Results}
\begin{figure}
\centering
\includegraphics[width=0.5\textwidth]{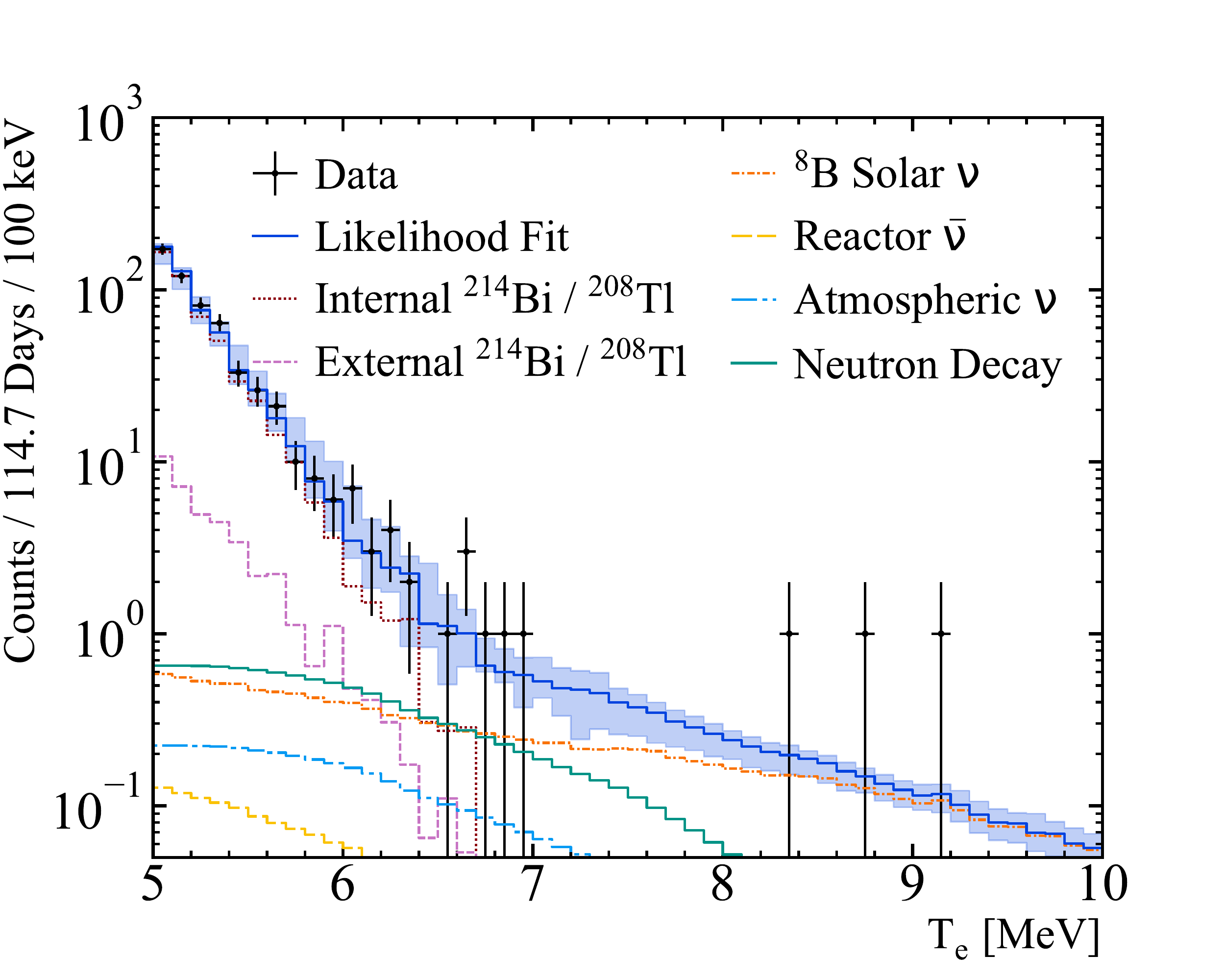}
\caption{Fitted energy spectrum across all data sets for neutron decay and
  backgrounds. Backgrounds were fit for each data set and the spectrum shown is
  the sum over all of the time-bins for the individual components. The
  contributions of Bi and Tl were fit independently for internal backgrounds but
  were merged in this plot. The signal is shown at its maximum likelihood, not
  at 90\% C.I. The errors around the full fit include the MC statistical
  uncertainties summed in quadrature with the individual systematic
  uncertainties. These errors are bin-by-bin correlated and are dominated by
  the energy resolution and the energy scale systematic uncertainties.}
  \label{fig:spectral_neutron}
\end{figure}

\begin{figure}
	\centering
	\includegraphics[width=0.5\textwidth]{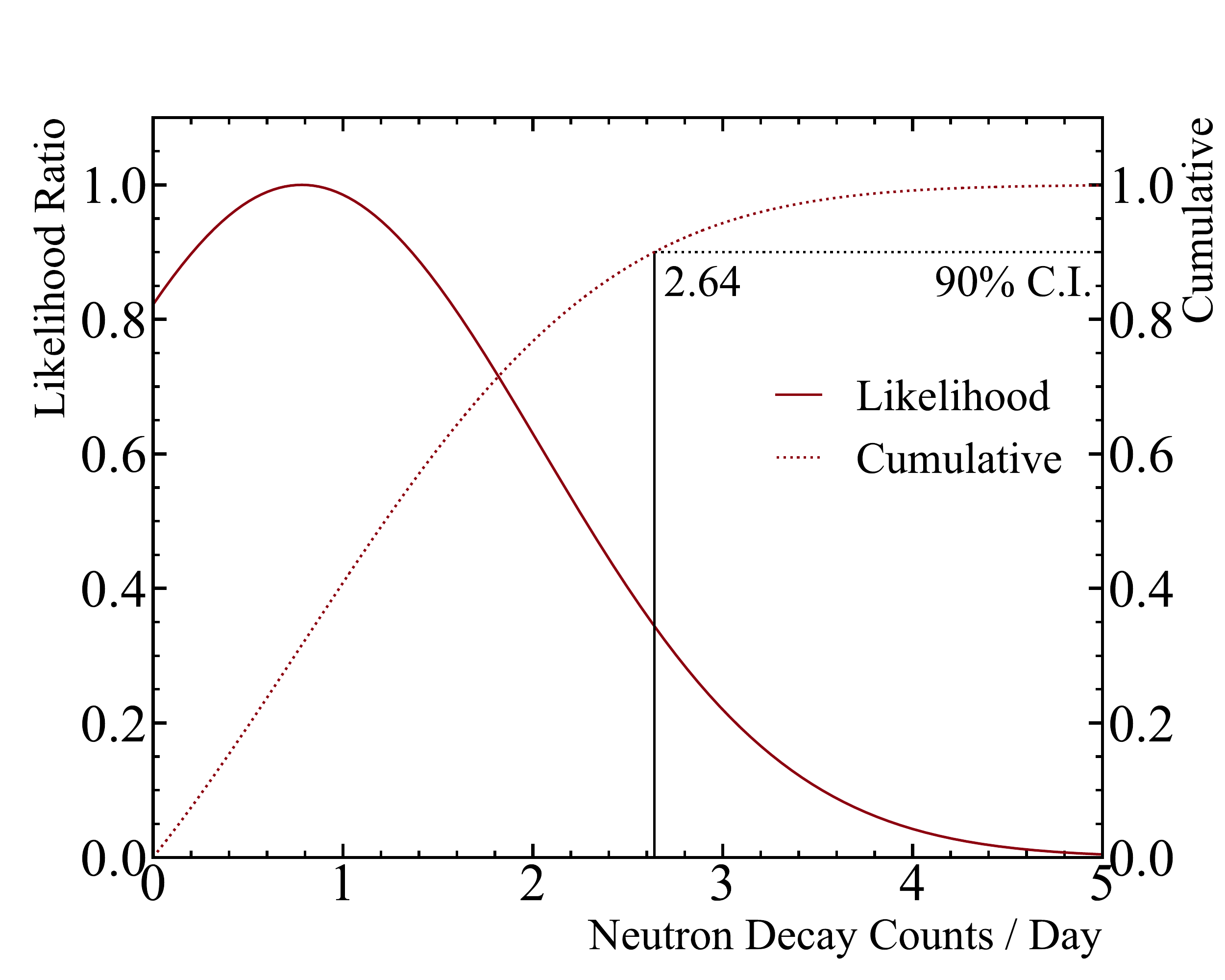}
        \caption{The likelihood ratio (the maximum likelihood as a function of
        the signal divided by the likelihood at the best fit value) for the
        neutron decay mode of the spectral analysis versus the limit of the
        number of signal decays per day. The associated cumulative distribution
        indicates the point at which the limit on the signal at 90\% C.I. is
        drawn.} \label{fig:SpectralLikelihood}
\end{figure}

The results of the spectral analysis for the neutron decay mode are shown in
Fig.~\ref{fig:spectral_neutron} with the fitted energy spectrum of the
neutron decay signal at its maximum likelihood value plotted alongside the
fitted backgrounds and data. Figure~\ref{fig:SpectralLikelihood} shows the
normalized and cumulative likelihood distribution for the neutron mode. The
resulting limits on each mode of invisible nucleon decay are shown in
Table~\ref{tbl:combinedresults} alongside the existing limits. A breakdown of systematic uncertainties is
given in Table~\ref{tbl:ndecay_spectral_systematics}.

\begin{table}
\label{tbl:combinedresults}
\centering
\begin{tabular}{c | c | c | c }
\hline \hline
& Spectral analysis & Counting analysis & Existing limits\\
\hline
$n $ & $2.5\times 10^{29}$~y & $2.6\times 10^{29}$~y & $5.8\times 10^{29}$~y~\cite{KamLANDNucleonDecay}\\
$p $ & $3.6\times 10^{29}$~y & $3.4\times 10^{29}$~y & $2.1\times 10^{29}$~y~\cite{SNONucleonDecay}    \\
$pp$ & $4.7\times 10^{28}$~y & $4.1\times 10^{28}$~y & $5.0\times 10^{25}$~y~\cite{Back:2003wj}        \\
$pn$ & $2.6\times 10^{28}$~y & $2.3\times 10^{28}$~y & $2.1\times 10^{25}$~y~\cite{Tretyak:2004ze}     \\
$nn$ & $1.3\times 10^{28}$~y & $0.6\times 10^{28}$~y & $1.4\times 10^{30}$~y~\cite{KamLANDNucleonDecay}\\
\hline \hline
\end{tabular}
\caption{Lifetime limits at 90\% C.I. for the spectral and counting analysis, including statistical and systematic uncertainties alongside the existing limits.}
\end{table}

\begin{table*}
\centering
\begin{tabular}{c|c|c|c|c|c}
\hline \hline
Systematic & $n$ (events/day) & $p$ (events/day) & $pp$ (events/day) & $pn$ (events/day) & $nn$ (events/day) \\
\hline
Best fit    & 0.66          & 0.55          & 0.57           & 0.99           & 2.34 \\
\hline
Energy scale & +0.42, $-0.21$ & +0.25, $-0.13$ & +0.21, $-0.12$ & +0.41, $-0.23$ & +0.53, $-0.28$\\
Energy resolution & $\pm1.01$ & $\pm0.67$ & $\pm0.60$ & $\pm1.11$ & $\pm1.20$\\
x-shift & $\pm 0.02$ & $\pm0.01$ & $\pm0.01$ & $\pm0.02$ & $\pm0.02$\\
y-shift & $\pm 0.01$ & $\pm0.01$ & $\pm0.01$ & $\pm0.02$ & $\pm0.03$\\
z-shift & +0.02, $-0.01$ & +0.01, $-0.01$ & +0.01, $-0.01$ & +0.03, $-0.01$ & +0.05, $-0.01$\\
xyz-scale & +0.14, $-0.13$ & +0.10, $-0.09$ & +0.10, $-0.08$  & +0.19, $-0.16$ & +0.31, $-0.25$\\
$\beta_{14}$ & $\pm 0.04$ & $\pm 0.03$ & $\pm 0.03$ & $\pm0.07$ & $\pm0.14$\\
Direction & +0.14, $-0.07$ & +0.11, $-0.07$ & +0.11, $-0.08$ & +0.21, $-0.13$ & +0.44, $-0.28$\\
\hline
Total (syst.) & +1.12, $-1.05$ & +0.73, $-0.69$ & +0.65, $-0.62$ & +1.22, $-1.15$ & +1.43, $-1.30$\\
Statistical & +0.57, $-0.48$ & +0.42, $-0.37$ & +0.42, $-0.40$ & +0.75, $-0.71$ & +2.16, $-1.59$ \\
\hline
90\% C.I. & 2.64 & 1.85 & 1.76 & 3.21 & 6.59 \\
\hline \hline
\end{tabular}
\caption{Systematic uncertainties and fit results for the spectral analysis.
  For each of the decay modes, the best fit is given as well as the difference
  between the best fit and the shift-and-refit value for each source of
  uncertainty. The total is the sum in quadrature of each of the separate
  systematic uncertainties assuming no correlation between the components. The
  statistical uncertainty, found by integrating the likelihood function prior
  to convolution with the systematic uncertainties using a Feldman-Cousins
  confidence interval to 1$\sigma$, is shown for comparison. Finally, the
  convolved likelihood function is integrated to 90\% to get the final limit
  shown on the last line.} 
  \label{tbl:ndecay_spectral_systematics}
\end{table*}

\begin{table}
\centering
\begin{tabular}{ c | c | c }
\hline
\hline
Data & Observed & Expected \\
set  & events   & events   \\
\hline
1 & 1 & 1.17$^{+4.60}_{-0.05}$ $^{+1.33}_{-0.39}$ \\
2 & 2 & 2.35$^{+4.62}_{-0.40}$ $^{+3.44}_{-0.81}$ \\
3 & 4 & 3.47$^{+4.60}_{-0.15}$ $^{+3.11}_{-0.96}$ \\
4 & 8 & 3.37$^{+4.60}_{-0.17}$ $^{+2.70}_{-0.98}$ \\
5 & 1 & 1.46$^{+4.60}_{-0.13}$ $^{+2.17}_{-0.60}$ \\
6 & 6 & 5.84$^{+7.40}_{-2.31}$ $^{+2.68}_{-0.62}$ \\
\hline
Total & 22 & 17.65$^{+12.68}_{-2.36}$ $^{+6.51}_{-1.85}$ \\
\hline
\hline
\end{tabular}
\caption{The observed and predicted number of events passing the counting
  analysis cuts for each data set. The uncertainty on the total number is found
  by treating each data set as independent and combining in quadrature. The
  first uncertainty is the statistical uncertainty, the second is the total
  systematic uncertainty.
	The statistical uncertainty is almost completely dominated by the background
	from the PMTs.}
  \label{tbl:ndecay_box_results}
\end{table}

\begin{table*}
\centering
\begin{tabular}{ c | c | c | c | c | c }
\hline
\hline
Systematic & $n$ (events/day) & $p$ (events/day) & $pp$ (events/day) & $pn$ (events/day) & $nn$ (events/day) \\
\hline
Energy resolution & 0.72   & 0.54 & 0.52 & 0.92 & 4.07\\
Energy scale    & 0.42   & 0.26 & 0.24 & 0.43 & 0.87\\
Position resolution & 0.11  & 0.09 & 0.09 & 0.16 & 0.63\\
Position shift   & 0.02   & 0.01 & 0.01 & 0.02 & 0.09\\
Position scale   & 0.03   & 0.03 & 0.03 & 0.05 & 0.18\\
Direction resolution & 0.03 & 0.03 & 0.03 & 0.05 & 0.19\\
$\beta_{14}$    & 0.04   & 0.03 & 0.03 & 0.06 & 0.24\\      
Trigger efficiency & 0.03  & 0.02 & 0.02 & 0.04 & 0.16\\
Instrumentals   & 0.04   & 0.03 & 0.04 & 0.06 & 0.25\\
\hline
Total systematic & 0.84   & 0.61 & 0.58 & 1.03 & 4.24\\
Statistics    & 0.30   & 0.24 & 0.25 & 0.44 & 1.81\\
\hline
\hline
\end{tabular}
\caption{Systematic uncertainties for the counting analysis on the upper
  limit (at 90\% C.I.) on the signal decays per day, shown as the difference to
  the unshifted value.} \label{tbl:ndecay_box_systematics}
\end{table*}

For the counting analysis, Table~\ref{tbl:ndecay_box_results} shows the
observed events compared to the predictions for each data set. The results of
the counting analysis are shown alongside those of the spectral analysis in
Table~\ref{tbl:combinedresults}. A breakdown of the systematic uncertainties is
shown in Table~\ref{tbl:ndecay_box_systematics}.

\section{Conclusion}
The results shown by the spectral and counting analyses in
Table~\ref{tbl:combinedresults} are in good agreement. In the case of the
dineutron decay mode, the spectral analysis performed significantly better due
to the difference in the spectral shape of the dineutron signal, which is not
taken into account within the counting analysis.

The limit set in this work on the lifetime of the proton decay mode of 3.6$\times 10^{29}$~y
is an improvement on the existing limit from SNO, however the
neutron mode limit of 2.5$\times 10^{29}$~y is weaker than the current limit
from KamLAND.

For the dinucleon modes, the $nn$ limit of 1.3$\times 10^{28}$~y does not improve upon the existing
limit set by KamLAND,
but the $pn$ and $pp$ mode limits of 2.6$\times 10^{28}$~y and 4.7$\times 10^{28}$~y
improve upon the existing limits by close to three orders of magnitude.

\begin{acknowledgements}
Capital construction funds for the SNO\raisebox{0.5ex}{\tiny\textbf{+}} experiment were provided by the Canada Foundation for Innovation (CFI) and matching partners. 
This research was supported by: 
{\bf Canada: }
Natural Sciences and Engineering Research Council, 
the Canadian Institute for Advanced Research (CIFAR), 
Queen's University at Kingston, 
Ontario Ministry of Research, Innovation and Science, 
 Alberta Science and Research Investments Program, 
National Research Council,
 Federal Economic Development Initiative for Northern Ontario,
Northern Ontario Heritage Fund Corporation,
Ontario Early Researcher Awards;
{\bf US: }
Department of Energy Office of Nuclear Physics, 
National Science Foundation, 
 the University of California, Berkeley, 
Department of Energy National Nuclear Security Administration through the Nuclear Science and Security Consortium; 
{\bf UK: }
Science and Technology Facilities Council (STFC),
the European Union's Seventh Framework Programme under the European Research Council (ERC) grant agreement,
the Marie Curie grant agreement;
{\bf Portugal: }
Funda\c{c}\~{a}o para a Ci\^{e}ncia e a Tecnologia (FCT-Portugal);
{\bf Germany: }
the Deutsche Forschungsgemeinschaft;
{\bf Mexico: }
DGAPA-UNAM and Consejo Nacional de Ciencia y Tecnolog\'{i}a.

We thank the SNO\raisebox{0.5ex}{\tiny\textbf{+}} technical staff for their strong contributions.  We would like to thank SNOLAB and its staff for support through underground space, logistical and technical services. SNOLAB operations are supported by CFI and the Province of Ontario Ministry of Research and Innovation, with underground access provided by Vale at the Creighton mine site.

This research was enabled in part by support provided by WestGRID (www.westgrid.ca) and Compute Canada (www.computecanada.ca) in particular computer systems and support from the University of Alberta (www.ualberta.ca) and from Simon Fraser University (www.sfu.ca) and by the GridPP Collaboration, in particular computer systems and support from Rutherford Appleton Laboratory~\cite{gridpp, gridpp2}. Additional high-performance computing was provided through the ``Illume'' cluster funded by CFI and Alberta Economic Development and Trade (EDT) and operated by ComputeCanada and the Savio computational cluster resource provided by the Berkeley Research Computing program at the University of California, Berkeley (supported by the UC Berkeley Chancellor, Vice Chancellor for Research, and Chief Information Officer). Additional long-term storage was provided by the Fermilab Scientific Computing Division. Fermilab is managed by Fermi Research Alliance, LLC (FRA) under Contract with the U.S. Department of Energy, Office of Science, Office of High Energy Physics.
\end{acknowledgements}

\bibliography{References}

\end{document}